\documentclass{aa}  
\usepackage{graphicx}
\usepackage{txfonts}
\def\simlt{\lower.5ex\hbox{$\; \buildrel < \over \sim \;$}}
\def\simgt{\lower.5ex\hbox{$\; \buildrel > \over \sim \;$}}
\def\simpropto{\lower.2ex\hbox{$\; \buildrel \propto \over \sim \;$}}

\begin{document}
   \title{Active galactic nuclei and  massive galaxy cores}

\author{S. Peirani\inst{1,}\inst{2}, S. Kay\inst{1,}\inst{3}, \and J. Silk\inst{1} }

    \institute{Department of Physics, University of Oxford, Denys Wilkinson Building,
 Keble Road, Oxford OX1 3RH, UK\\
              \email{sp@astro.ox.ac.uk ;
	 	 silk@astro.ox.ac.uk}
        \and
	 Institut d'Astrophysique de Paris, 98 bis Bd Arago, 75014 Paris, France - Unit\'e mixte de 
	 recherche 7095 CNRS - Universit\'e Pierre et Marie Curie 
	\and
	Jodrell Bank Centre for Astrophysics,
	Alan Turing Building,
	School of Physics and Astronomy,
	The University of Manchester,
	Manchester M13 9PL, UK \\
        \email{Scott.Kay@manchester.ac.uk}
             }

   \date{Received ..., ...; accepted ..., ...}

 
  \abstract
   {Central active galactic nuclei (AGN) are supposed to play a key role in
    the evolution of their host galaxies. In particular, 
    the dynamical and physical properties of the gas core must be affected by
    the injected energy.}
   {Our aim is to study the effects of
an AGN on the dark matter profile and  on the central stellar light distribution in
massive early type galaxies.}
   {By performing self-consistent N-body simulations,
 we assume in our analysis that periodic bipolar outbursts from a central AGN can induce 
harmonic oscillatory motions on both sides of the gas core.}
   {Using  realistic AGN 
properties, we find that the motions of the gas core, driven by such feedback processes,
can flatten the dark matter and/or stellar profiles after 4-5 Gyr. 
 These results are consistent with recent observational studies
 that suggest that most  giant elliptical galaxies
 have  cores or are ``missing light'' in their
inner part. Since stars behave as a ``collisionless''
 fluid similar to dark matter, the  density profile both  of stars  and dark matter should
 be affected in a similar way, 
leading to an effective reduction in the central brightness. }
   {}


   \maketitle
%

\section{Introduction}

 The cold dark matter (CDM) paradigm (Cole et al. \cite{cole}
and references therein) has led to a successful explanation of  the 
large-scale structure in the galaxy distribution on scales 0.02 $\simlt k \simlt$ 0.15h Mpc$^{-1}$.
The CDM power spectrum on these scales derived from 
large redshift surveys
 such as, for instance, the Anglo-Australian 2-degree Field 
Galaxy Redshift Survey (2dFGRS), 
is also consistent with the Lyman-$\alpha$ forest data in the 
redshift range 2$< z < $4 (Croft et al. \cite{croft}).

In spite of these impressive successes, there are still discrepancies between simulations and 
observations on scales $\simlt$ 1 Mpc. We may mention 
 the large number of 
sub-$L_\ast$ subhalos present in 
simulations but not observed 
(Kauffmann, White \& Guiderdoni \cite{kauffmann}; Moore et al. \cite{moore1};
Klypin et al. \cite{klypin1}; but see 
Belokurov et al. \cite{belokurov} for an observational update), 
and the  excess of massive early-type galaxies undergoing ``top-down" assembly with high
 inferred specific star formation rates relative to
predictions of the hierarchical scenario (Glazebrook et al. \cite{glazebrook};
Cimatti, Daddi \& Renzini \cite{cimatti}).
Here we examine 
the sharp central density
cusp predicted by simulations in dark matter halos and not seen in the rotation 
curves of low surface brightness  galaxies (de Blok et al. \cite{blok}) or in bright
spiral galaxies (Palunas \& Williams \cite{palunas}; Salucci \& Burkert \cite{salucci}, 
Gentile \cite{gentile}).  Little attention has been
given to the corresponding situation in massive early-type galaxies, the focus of the present analysis.

Our model is presented as a toy model to justify an in depth study. It is in the same spirit as that developed by Mashchenko, Couchman \& Wadsley (\cite{mashchenko}) to account for cores in dwarf galaxies. Indeed it has a similar problem: where has the gas gone? But we stress that there are many ways of accounting for this and other objections (we provide  more details below).
The point we wish to make is that central cores in massive ellipticals are possible tracers of violent AGN activity, and provide an alternative to binary black hole scouring. If AGN are powerful sources of negative feedback that cleans out the gas supply to avoid forming excessively luminous and blue ellipticals, our model
is a plausible and even inevitable precursor. These cores are seen in the stellar component, and are likely present in the underlying (although subdominant) dark matter distribution.

Navarro, Frenk \& White (\cite{nfw1}, \cite{nfw2})  have shown that the 
spherically-averaged density profiles of simulated halos can be fitted by a
 simple analytical function, depending
on a characteristic density $\rho_*$ and a characteristic radius $r_s$:
\begin{equation}
\rho_{dm}(r)=\frac{\rho_*}{r/r_s(1+r/r_s)^2} \;.
\end{equation} 
This profile (dubbed the NFW-profile) is steeper than 
that of an isothermal sphere at large radii and shallower close to the center. 
 Steeper 
profiles in the central regions ($\alpha \sim$ -1.5) have been found in high resolution simulations  
(Moore et al. \cite{moore2}; Ghigna et al. \cite{ghigna}; Fukushige \& Makino \cite{fukushige}). 

While the ``universality"  of density profiles of dark halos is still a matter of 
debate, possibly depending on the merger history (Klypin et al. \cite{klypin2}; Ricotti 
\cite{ricotti}; Boylan-Kolchin \& Ma \cite{boylan}) or on the initial conditions (Ascasibar et al. \cite{ascasibar}),
it could nevertheless be an important key to understanding the mechanism(s) by which these 
systems relax and attain equilibrium. In a collisionless self-gravitating system, collective 
mechanisms such as violent relaxation and/or phase mixing  certainly play a major role
in the relaxation of halos, but density profiles may also be affected by 
gravitational scattering of dark matter particles in substructures present inside halos
(Ma \& Bertschinger \cite{ma_bert}).
In fact, recent numerical experiments indicate that this diffusion process may alter 
significantly the inner density profile of halos, producing a flattening of the original 
profile within a few dynamical time scales (Ma \& Boylan-Kolchin \cite{ma_boylan}). 
The most detailed study (Ricotti \cite{ricotti})
of N-body simulations at different scales concluded that galaxies have shallower
density profiles  than clusters.

Central cusps seen in simulations could also be understood as a consequence of 
the inflow of low-entropy material and therefore contain information on the relic 
entropy of dark matter particles. In fact, high resolution 
simulations of galaxy-sized CDM halos indicate an {\it increase} of the coarse-grained phase-space
density Q (defined as the ratio between the density and the cube of 1-D velocity 
dispersion in a given volume, e.g., $Q = \rho/\sigma^3$ ) towards the center (Taylor \& Navarro \cite{taylor}).
Similar results were obtained by Rasia, Tormen \& Moscardini (\cite{rasia}), who also obtained a 
power-law variation, {\it e.g.}, $Q \propto r^{-\beta}$ for cluster-size halos, with $\beta$ quite
close to the value found by Taylor \& Navarro (\cite{taylor}), namely, $\beta \approx$ 1.87.

Several solutions have been proposed to explain such discrepancies between observations and
numerical simulations. For example, the dark matter (DM) can be ``heated" by the baryons
by dynamical friction due to
self-gravitating gas clouds orbiting near the center of the galaxy (El-Zant, Shlosman \& Hoffman \cite{el-zant}, El-Zant et al. \cite{el-zant2}),
by the evolution of a stellar bar (Weinberg \& Katz \cite{weinberg};
Holley-Bockelmann, Weinberg \& Katz \cite{holley-Bockelmann};
Sellwood \cite{sellwood}), by
the radiation recoil by a black hole (Merritt et al. \cite{merrit}) or by
random bulk gas motions driven by supernovae feedback, recently suggested by
Mashchenko, Couchman \& Wadsley (\cite{mashchenko}). Other mechanisms have been proposed such as the
the transfer of angular momentum from baryonic to dark matter (Tonini, Lapi \& Salucci\cite{tonini})
or the expulsion of a large fraction of the gas due to feedback activities, causing the dark matter to expand (Gnedin \& Zhao \cite{gnedin}).

Most of the previous studies  consider the scales of dwarf galaxies or those of
galaxy clusters in order to interpret the observed DM  profiles in these objects
which are thought to be dark matter dominated in their inner regions.
In the present paper, we focus our attention on the scale of a typical giant elliptical galaxy,
and we investigate the effects of a central AGN in its core. Similar but more
 extreme feedback ideas have been proposed for disk galaxies such as the Milky Way,
where massive early outflows were invoked that resulted in homogenizing and
 heating the inner dark halo (Binney, Gerhard \& Silk \cite{binney}).

Black holes and
associated AGN feedback are generally considered
to be active in the centers of giant galaxies. They must represent  huge sources of 
energy and therefore are thought to play a key role in the formation of bright ellipticals (Silk \cite{silk}).
They can also be responsible for heating the gas core to regulate the cooling
flows in both  elliptical galaxies (Best et al. \cite{best1}) and brightest group and cluster galaxies
(Dunn \& Fabian \cite{dunn}; Best et al. \cite{best2}). 
Although over the past few years, important progress have been done
with high spectral resolution using HST, Fuse, Chandra and XMM-Newton, the nature of the interaction between the AGNs and the surrounding intergalactic medium is still poorly understood. However, observations
from UV and X-ray absorption lines of bright AGNs seem to suggest their outbursts can induce sonic motions to gas clouds and lead to important biconical mass outflows (see for instance Hutchings et al. \cite{hutchings};
Bridges \& Irwin \cite{bridges}; Churazov et al. \cite{churazov} and references therein).  
In the present work, we assume
 that each AGN outburst can induce a bulk
motion to the gas core by energy transfer. We then  explore the efficiency of
 this mechanism of gas core bulk motions 
for flattening the central DM  and stellar cusp.
This paper is organized as follows:  in section 2, we
describe our toy model, we present the results in section 3, and in section 4, our main conclusions are
summarized.

\section{Toy model }

We first develop a toy model which consists of building
an N-body realization of an isolated, equilibrium model galaxy.
A direct method uses the distribution function (DF),  which provides the relative probability
 of a star to have a certain position and velocity, and then 
use a Monte-Carlo sampling of the DF to generate the N-body realization.
Thus, assuming spherical symmetry, we generate a spherical equilibrium DM halo
 with a Hernquist density profile (Hernquist \cite{hernquist}):
\begin{equation}
\rho_{dm}(r)=\frac{M_{dm}}{2\pi}\frac{a}{r(r+a)^3}
\end{equation} 
where $M_{dm}$ is the total mass of dark matter and $a$ the scale radius. In our
fiducial model, we use $M_{dm}=10^{13}\,h^{-1}M_\odot$ and $a=77.0\,h^{-1}$ kpc.
For comparison, this profile is identical in the inner parts to an NFW-profile
with  virial radius $r_{200}=445\,h^{-1}$ kpc ($r_{200}$ defines the sphere within which the mean density
is equal to 200 times the critical density) and scale radius $r_s=44.5\,h^{-1}$ kpc respectively
(see for instance relation 2 between $a$ and $r_s$ in
Springel, Di Matteo \& Hernquist \cite{springel1}). These 2 profiles
are compared in figure \ref{fig1}.
The choice of using an Hernquist profile is motivated by two main reasons.
Firstly, the density declines faster than an NFW-profile in the outer parts of the halo. Thus, for
an isolated halo, truncation 
at the virial radius is not needed. Furthermore, the Hernquist profile has an analytical
expression for both the distribution function and the velocity dispersion
 (Hernquist \cite{hernquist}), which is not the case for an NFW-profile.
By proceeding in this way, we try to minimize undesirable effects which can lead to unstable equilibrium.
 Conventionally, the scale radius is given in 
terms of the concentration parameter $c=r_v/r_s$. For our fiducial model, we use
$c=10,$ in good agreement with values found in previous cosmological N-body simulations
(Bullock et al. \cite{bullock1}; Dolag et al. \cite{dolag}).

The gas component follows a Hernquist profile as well with a total gas mass
$M_{gas}$ constrained by  the baryonic fraction $f_b=M_{gas}/(M_{dm}+M_{gas})$.
Its initial temperature profile is chosen
so that the gas is initially in hydrostatic equilibrium.

To model the effects of the central AGN on the gas component, we use a simple
representation where the outflow takes the form of a 
collimated jet which causes the gas to expand by compression. Since the resulting detailed shape of the gas is complicated and poorly known, we use
 the following simplified model of spherical symmetry (spherical symmetry makes the model well-defined with the minimum number of free parameters).   
 We first  define a gas core
 radius which limits the region affected by the AGN activity. We assume
its value to be $30\,h^{-1}$ kpc corresponding to less than 10\% of the virial radius
and includes a  few percent of the total mass of the system.
Then the gas core is split into 2 parts and due to periodic AGN outbursts,
each part has an independent harmonical motion on opposite sides 
of the AGN. 
The frequency and the amplitude of the oscillations depend on the frequency of the AGN activity,
on the energy transfer between the AGN and the gas, and on the 
physical properties of the gas (viscosity, temperature, local density, etc...). To simplify matters, we  
assume that i) the frequency of the AGN emission matches the frequency of the oscillations of the gas
and ii) the amplitude of motions of a specific model corresponds to a fractional
 displacement of the scale radius. 
Finally, we have also studied, in a second experiment, 
how our results are affected when a stellar component is
included. 

\section{Results}   
\subsection{Adiabatic gas}

We carry out each simulation with the public code GADGET2 (Springel \cite{springel2}) where $5\times 10^5$
dark matter particles and $5\times 10^5$ gas particles have been used. 
The value of the softening length is  $1\,h^{-1}$ kpc for all experiments and
the baryonic fraction is taken to  $f_b=0.13$. Although the gas is supposed
to have a central core, our representation of a  Hernquist profile will not affect our results since
the relevant parameter is the total  mass of the core gas.
In our study, its resulting value is taken to be $1.2\times 10^{11} M_\odot$, i.e. 1\%
 of the total mass of the system. 

\begin{figure}
\rotatebox{0}{\includegraphics[width=\columnwidth]{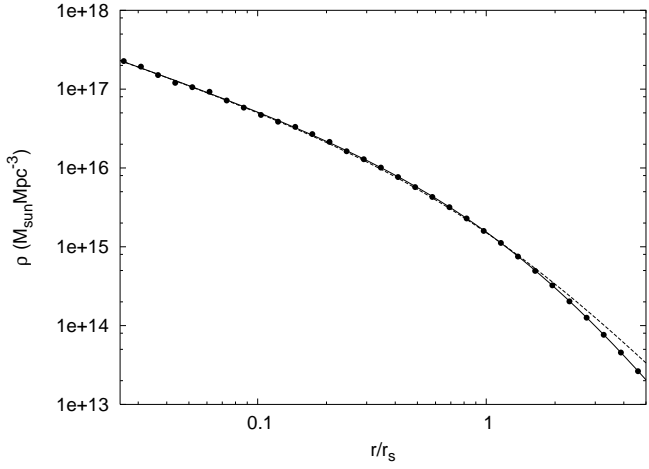}}
\caption{DM density profile of our unperturbed model. The dots
are estimates of the density at $5\,h^{-1}$ Gyr while the solid line 
is the expected Hernquist profile whose relevant parameters are discussed
in the text. For comparison,
the ``equivalent'' NFW-profile is also shown (dashed line).}
 \label{fig1}
 \end{figure}

The strongest effects are expected to occur when the velocities of the gas particles
are close to  the velocity dispersion of dark matter. 
Then, in our fiducial model, we consider that each injection of energy by the AGN produces a gas bulk motion 
 with an initial velocity of $V_{gas}=260$ km/s, corresponding
 to the velocity dispersion of
dark matter at radius $10-20\,h^{-1}$ kpc. Since the escape velocity at  $r=30\,h^{-1}$ kpc is
about 900 km/s, this ensures that the two moving parts of the gas core are always bound to the system.
Then, we let the system evolve
during $5\, h^{-1}$ Gyr which corresponds to about 20 times the dynamical time of the gas core.
Figure \ref{fig1} shows the DM density profile at $t=5 h^{-1}$ Gyr for an unperturbed system ($V_{gas}=0$).
The whole system seems to be in a very stable equilibrium since the estimated densities at different radii 
match well  those of the initial density profile.

\begin{figure}
\rotatebox{0}{\includegraphics[width=\columnwidth]{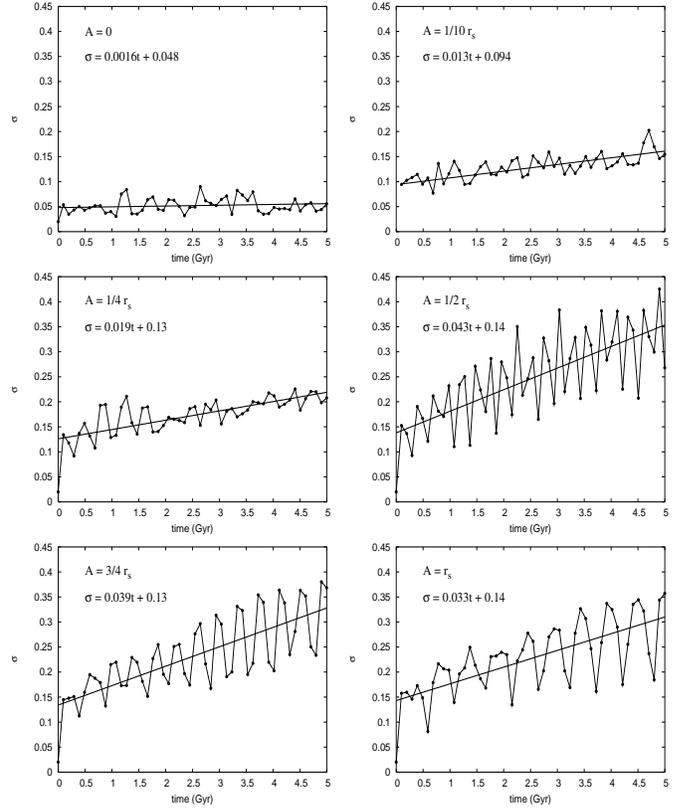}}
\caption{Time evolution of the parameter $\sigma$ for the model
with $V_{gas}=260$ km/s and for 6 different amplitudes A.
For each experiment, the solid line is the best fit to the data. }
 \label{fig2}
 \end{figure}

To study the time evolution of the degree of cuspiness of each model, we use a simple estimator defined by
\begin{equation}
\sigma=\sqrt{\frac{1}{N}\sum_i^N\Big(\frac{\rho_{exp}(r_i)}{\rho_{theo}(r_i)}-1\Big)^2}
\end{equation} 
\noindent
where $\rho_{exp}(r_i)$ is the estimated density at the radius $r_i<50\,h^{-1}$ kpc
 compared to  the theoretical
value $\rho_{theo}(r_i)$. $N$ represents the number of measures used for the estimation
 of $\sigma$. Typically, we took
$N=50$ points separated within logarithmic bins of equal width.
 It is worth mentioning than for each time-step we have estimated the mass center
 of the DM component by using a friend-of-friends algorithm (Davis et al. \cite{davis}).
The value of the linking length  is 0.05 in units of  the mean interparticle separation,
 corresponding to evaluating the center of mass of the 5000 DM particles.  
Figure \ref{fig2} shows the evolution of our estimator $\sigma$ for 6 specific amplitudes $A$ of the
oscillations. First, when the system is at rest, namely $V=0$ and $A=0$, $\sigma$ is constant and close to 0
as expected. This experiment
is used to check the stability of the equilibrium of our system and to estimate the maximum numerical error. 
The positive values are mostly due
to the lack of resolution in the very inner parts of the halo ($r_i<0.03\, r_s$) which can lead to
underestimates or overestimates of the density.
For both $A=1/10\,r_s$ and $A=1/5\,r_s$, in spite of erratic fluctuations, on average, a small
 increase of $\sigma$ is observed.
On the contrary, for $A = 1/2\,r_s$, $3/4\,r_s$ and $r_s$,   $\sigma$ increases faster
 but no significant differences can been seen between the 3 scenarios. 
 
 However, two important remarks can be made.
Firstly, the strongest effects  occur when $A=1/2r_s$ where the DM particles seem to
enter into resonance. Secondly, the value of $\sigma$  oscillates
with the same frequency as the gas oscillations. This phenomenon can be clearly seen
 for $A=r_s$ where the period of
the oscillations is $0.52\,h^{-1}$ Gyr. This suggests that the shape of the inner
 profile continuously evolves
over  time  through different phases.

\begin{figure}
\rotatebox{0}{\includegraphics[width=\columnwidth]{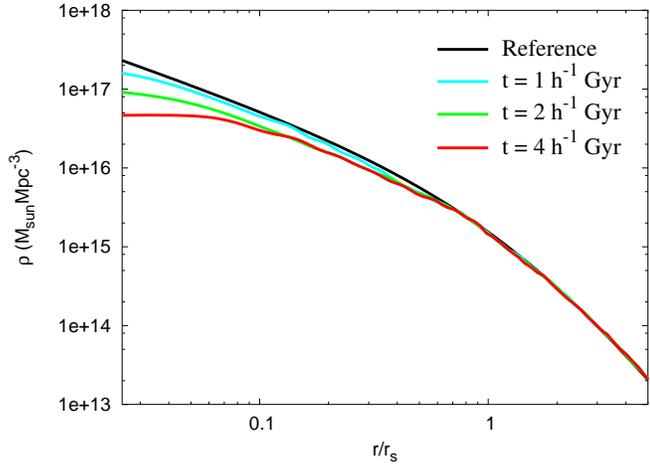}}
\caption{ DM density profile for $V_{gas}=260$ km/s and $A=1/2\,r_s$ plotted
at three different times, $t=1$ (dashed line), $t=2$ (dotted line) and $t=4\, h^{-1}$ Gyr.
 The solid line represents
the expected profile in the absence of any perturbation.}
 \label{fig3}
 \end{figure}

Next, we consider the case $A=1/2\,r_s$.
Figure \ref{fig3} represents the evolution of the DM density profile at $t=1,2$ and $4\, h^{-1}$ Gyr.
We notice that the DM cusp is progressively flatten. At $t=4\, h^{-1}$ Gyr, the DM profile can be  decomposed  
into a flat central part ($r<0.1 r_s \sim 5 h^{-1} kpc$) and a steeper outer part until it converges to the
original slope at $r\sim 0.7 r_s$. In this case,  
the derived density profile   cannot be well fitted by a Burkert profile (Burkert \cite{burkert}).

\begin{figure}
\rotatebox{0}{\includegraphics[width=\columnwidth]{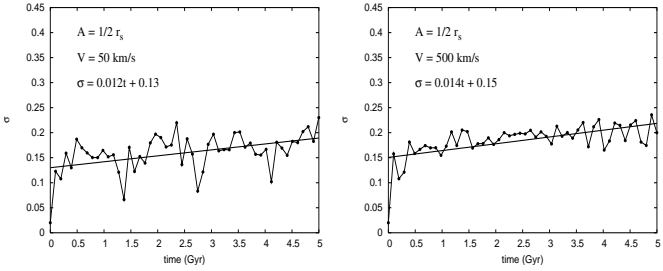}}
\caption{Time evolution of the parameter $\sigma$ for $A=1/2\,r_s$ and
for $V_{gas}=50$ km/s (right panel) or $V_{gas}=500$ km/s (right panel). }
 \label{fig4}
 \end{figure}


Figure \ref{fig4} shows the evolution of
the time evolution of
$\sigma$ when the velocity of the
gas is either small ($V_{gas}=50$ km/s) or high ($V_{gas}=500$ km/s) compared to the velocity dispersion
of the dark matter. In both case, an increase of $\sigma$ is observed but the 
effects are quite small, in good agreement with what we expect.
It is  interesting to notice that for $V_{gas}=50$ km/s, the three peaks correspond to 
1, 2 and 3 oscillation periods respectively (the period
of the oscillation is about $1.4\,h^{-1}$ Gyr). At these specific times, the effects on the
DM profile are negligible.

\subsection{Including a stellar distribution}

One way to improve our model is to include the formation of the elliptical
galaxy itself; dissipation of the gas (from radiative cooling processes) and 
subsequent star formation leads to a steeper initial DM density profile due
to adiabatic contraction. We achieve this by building new {\it initial} conditions
resulting from the merger of 4 sub-haloes; this is done in order to boost the star 
formation rate (SFR) and form a giant galaxy in a relatively short period. 

Each sub-halo has a mass of $2.5\times 10^{12} M_{\odot}$ and the same density profiles
as discussed in section 2. Additionally, we give each sub-halo angular momentum by
setting the spin parameter, $\lambda \equiv J|E|^{1/2}/GM^{5/2}\approx 0.1$ 
(where $J$ is the angular momentum, $E$ is the total energy of the halo and $M$ is its mass),
and the specific angular momentum distribution, $j(r) \propto r$, as expected from 
cosmological $N$-body simulations (Bullock et al. \cite{bullock2}). The 4 respective centers of mass 
are initially placed at rest in a cross configuration, of length $400\; h^{-1}$ kpc.
Each opposing pair has oppositely orientated angular momentum vectors in order to avoid
an excessive amount of rotation after the merger event. 

Before repeating our previous experiment, we first let the above configuration evolve
for 6.5 $h^{-1}$ Gyr. During the first 1.5 $h^{-1}$ Gyr (which captures the merger event), we turn on 
radiative cooling and star formation, then allow the system to relax for the remaining
time. Cooling and star formation were introduced in the manner of Katz, Weinberg \& Hernquist \cite{katz}.
The cooling prescription allowed each gas particle with $T>10^4$K to cool at
constant density for the duration of each timestep, assuming a metallicity, $Z=0$.
Each gas particle with $T< 2\times 10^4 K$ was then deemed eligible for star formation
and we adopted the usual SFR prescription
$\frac{d\rho_*}{dt} = c_* \frac{\rho_{gas}}{t_{dyn}}$,
where $\rho_*$ and $\rho_{gas}$ refer to the stellar and gas density respectively, $t_{dyn}$ is the
dynamical time of the gas and $c_*=0.5$ the star formation efficiency (this value is quite high to
boost the SFR). Rather than spawning new (lighter) star particles, we implemented the above
prescription in a probabilistic fashion. Assuming a constant dynamical time across the timestep,
the fractional change in stellar density, $\Delta \rho_*/\rho_* = 1-\exp(-c_* \Delta t/t_{\rm dyn})$.
We drew a random number, $r$, from the unit interval and converted a gas particle to a star if
$r<\Delta \rho_*/\rho_*$.

After 1.5 $h^{-1}$ Gyr, a galaxy with stellar mass, $M_*=4.5\times 10^{11} M_{\odot}$ formed with 75 per cent of 
the stars being born within the first Gyr.
During the next 5 $h^{-1}$ Gyr, we let the system relax to make sure that our
new initial conditions are stable.
  It is worth mentioning that our aim here is not to perform a 
detailed study of the formation of an elliptical galaxy (which is poorly understood), but to construct
a sensible starting point in order to assess the gross effect of forming the galaxy (i.e. from 
dissipation) on our previous results, as well as the effect on the stellar distribution itself, which
like the dark matter, is a collisionless fluid.

From our new initial conditions, we perform the same simulation and analysis as described in the previous
section. We use here $A=1/2\,r_s$ and $V_{gas}=320$ km/s corresponding to the average of the
initial velocity dispersion of dark matter in the inner part. We justify shutting off cooling and star
formation by assuming that the AGN activity balances cooling in our relaxed system, as suggested by 
Best (\cite{best1}). Although a significant fraction
of gas has been converted into stars, the mass of the gas at $30\,h^{-1}$ kpc
is still substantial ($10^{11} M_{\odot}$) due to the subsequent inflow of 
hot gas from the outer part of the halo. 

\begin{figure}
\rotatebox{0}{\includegraphics[width=\columnwidth]{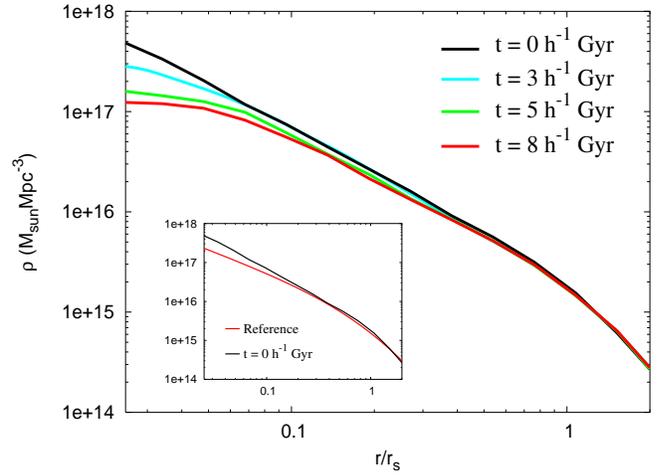}}
\caption{DM density profile for $V_{gas}=320$ km/s and $A=1/2\,r_s$ plotted
at three different times, $t=3$ (cyan), $t=5$ (green) and $t=8\, h^{-1}$ Gyr (red).
 The black lines represent the initial DM profile (with a stellar component)
  which is steeper than our fiducial
 model discussed in section 2 and represented in red in the small panel. }
 \label{fig5}
 \end{figure}

\begin{figure}
\rotatebox{0}{\includegraphics[width=\columnwidth]{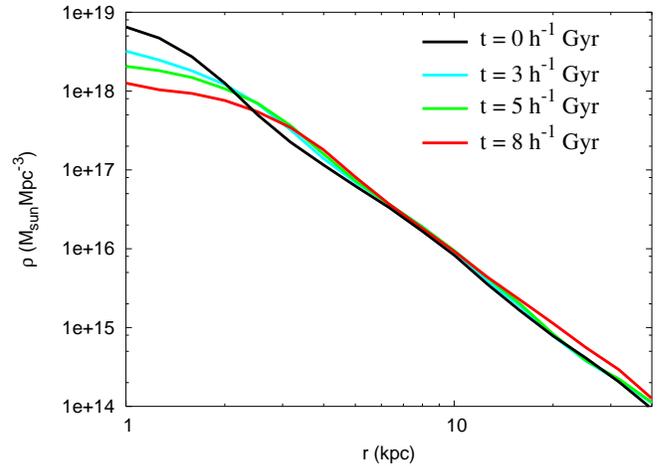}}
\caption{Stellar density profile for $V_{gas}=320$ km/s and $A=1/2\,r_s$ plotted
at $t=0$ (black), $t=3$ (cyan), $t=5$ (green) and $t=8\, h^{-1}$ Gyr (red).. }
 \label{fig6}
 \end{figure}

Figure \ref{fig5} shows the evolution of the DM density profile at $t=0,3,5$, and $8\, h^{-1}$ Gyr.
The DM cusp is again progressively flattened but a longer time is needed to obtain an effect of 
similar magnitude to the model with no stars. 
However, no significant variation is found between 5 and 8 $h^{-1}$
Gyr suggesting that most of the effect happens during the first 5 $h^{-1}$ Gyr.
It is also interesting to note that a similar effect
is found with the stellar density profile, as shown in Fig. \ref{fig6}. In particular, we found a 
significant deficit of stars within $2\; h^{-1}$ kpc. Here again, noticeable effects are found
during the first 5 $h^{-1}$ Gyr.

Before drawing any conclusions, we need to check if other physical mechanisms
are susceptible to play a role in the flattening of the DM profile. In 
particular, after the phase of galaxy formation from baryonic collapse and/or
sub-haloes merger, residual clumps of gas are expected to
exist and move through the virialized regions of the halo. If these structures are much more massive than DM particles and survive long enough from
tidal interactions, then their orbital energy can be transfered to DM
through dynamical friction.
Consequently, the distribution of DM should be affected, 
especially in the inner part.
This effect was clearly identified and characterized by
El-Zant, Shlosman \& Hoffman (\cite{el-zant}) and El-Zant et al. (\cite{el-zant2})
for both galaxy  and
clusters of galaxies. 
They found that the dynamical friction induced on clumps of gas with at least a mass
of 0.01 percents of the total mass of the system can flatten the DM cusp after a few Gyr.

In order to investigate if the mechanism proposed by El-Zant et al. (2001, 2004) 
is operating in our model, we have
measured the clumpiness of gas present in our initial conditions. To do that, 
potential sub-structures have been identified by 
a friend-of-friend algorithm. We used different values of the linking length
and, for each of them, we have estimated the number of sub-structures by using
the following procedure. First,
once a structure is identified, the total energy of each 
particle is computed, with respect to the centre of mass, and
those with positive energy are removed.
The procedure is repeated with the new  centre of mass, computed according
to its usual definition, until no unbound particles are found.
Then,  only structures with at least $57$ 
particles of gas, corresponding to objects with a mass greater than 10 times
the DM particle mass (or 0.002 percents of the total mass of the system), were retained.

We found that whatever the linking length, only one bound structure satisfying
the above criteria is found at most (i.e the center
of the halo itself). This proves that no bound sub-structures exist 
and that the gas is distributed in a smooth way (similar result is found for the stellar component). 
This is not surprising since we used a simplified prescription for
the star formation in which  the efficiency was quite high and
no standard criterion on the overdensity of the gas has been taken into account.
Moreover, we let the whole system relax during 5 $h^{-1}$ Gyr after the merger event.
Therefore, the lack of clumpiness of gas in our numerical model
allows us to ignore the influence of dynamical friction.
However, there is no denying that in reality,
inhomogeneities and clumpiness of gas are expected to be present and so the dynamical friction acting on them may not be negligible.
 Assessing which mechanism will dominate is beyond the scope of the present paper as our aim was to provide a plausible alternative mechanism,
 however we note that the effect found by El-Zant et al. (\cite{el-zant2}) at small fractions of the scale radius and after a period of 4 Gyr, seems to have a similar amplitude to our result.
We can therefore reasonably think that some combination of dynamical friction acting on baryonic clumps and the mechanism proposed in this paper would significantly affect the slope of the inner DM profile.

\section{Discussion and Conclusions}

More and more observations indicate that the DM profiles of dwarf galaxies have
a central DM core, in contradiction with results of numerical simulations. 
On the contrary, clusters tend to have a central cusp suggested
by recent studies combining strong and weak lensing (see for instance, Limousin et al. \cite{limousin};
Leonard et al. \cite{leonard}; Umetsu, Takada, \& Broadhurst \cite{umetsu}) even exception can be found
(Sand, Treu \& Ellis \cite{sand}).
In this paper, we focus our attention on intermediate scales corresponding in particular to massive
 early-type galaxies.
These are known to contain a central SMBH, whose mass scales with the spheroid velocity dispersion.
  The growth phase of this SMBH must have been nearly contemporaneous with the epoch of spheroid star
 formation. Which came first is controversial and not known,  but it is eminently plausible that
 strong  AGN activity occurs when the bulk of the spheroid stars have formed. For example, the
 epoch of activity of 
AGN as studied via hard x-rays yields a comoving growth rate that  peaks at  the same epoch as
 that of the cosmic star formation rate
at $z\sim 2$. The most massive spheroids are most likely already in place.
 Nevertheless, the presence of SMBHs of mass $\sim 10^9$ $M_\odot$ at high redshift ($z\sim6$)
   presents some challenges to this theoretical model since it's not clear at all how can
   such massive objects and their host halos (with probably a mass greater than $10^{12} M_\odot$) can form so early in a $\Lambda$CDM cosmology.
    One possible solution is that intermediate mass  black holes
    form as remnants of Population III stars in the first generation of clouds (Spaans \& Silk \cite{spaans}).
   Then, accretion is the preferred
   mechanism given that the quasar luminosity function matches the local BH mass function for plausible
   accretion efficiency (Yu \& Tremaine \cite{yu}). As far as the most massives halos being
   in place, yes, if they are rare as seems to be the case at high redshift. 
   It's interesting to note that recent cosmological simulations from Li et al. \cite{li} of formation of high redshift quasar from hierarchical galaxy
   mergers seems to favor this scenario.

Using N-body equilibrium systems, we have used a simple description of the AGN activity
in which the energy transfer generates a harmonic bulk motion on both sides of the gas core. We then
investigate the impact of this mechanism on the  DM core profile. 
For a typical giant galaxy of total mass $M\sim 10^{13} M_\odot$, we find that particularly
 strong effects occur when  
i) the initial velocity
of the gas bulk motion is close
 to the velocity dispersion of the dark matter
 and ii) when the amplitude of the oscillations of the bulk motions extends over  half the scale radius. 
In this case, the DM cusp  progressively flattens after $4-5 h^{-1}$ Gyr (with or without a
stellar distribution) but
cannot be well fitted by a Burkert profile.
However, the effects obtained are not negligible
and  our  model can be easily improved. 
For example, 
One could imagine that the direction and
the efficiency of the outbursts do not  remain constant in time.
On the other hand, if we believe that ellipticals form by major mergers, 
then these merger events may lead to formation of black hole binaries. Binaries provide
 an alternative DM and stellar core  heating mechanism. In this case, the combination of two
AGN feedback  modes may enhance the observed effects.
Finally, the combination with other physical mechanisms may also improve the scenario.
In particular, the dynamical friction acting on clumps of gas
is expected to have a strong contribution, as shown by
 El-Zant, Shlosman \& Hoffman (\cite{el-zant}) and El-Zant et al. (\cite{el-zant2}).
 Taking into account this process, the effects obtained in the paper should be amplified.

Our fiducial model is not unreasonable. For instance, 
Voit \& Donahue (\cite{voit})  claim that AGN outbursts of about $10^{45}$ erg/s,
 lasting for at least $10^7$ yr and occurring every $10^8$ yr
are required  in order to explain the inner entropy profile in cooling-flow clusters.
 This equates to an energy outburst of
$10^{59}$ erg, comparable to the total kinetic energy periodically injected into the gas core, assuming
   $V_{gas}=260-320$ km/s. The frequency
of the AGN emission is
also in good agreement with the frequency $\nu$ of the gas motions since
$\nu \sim2.6\times 10^8\,h^{-1}$ yr and  $\nu \sim2.2\times 10^8\,h^{-1}$ yr
respectively for $V_{gas}=260$ km/s and $V_{gas}=320$ km/s, using $A=1/2\,r_s$. 
 Furthermore, it's worth mentioning that in the present case, the dynamical time of the core gas
is comparable to the time between outbursts, 
taking a mean density of gas of $\sim 5\times 10^{15} M_{sun}.Mpc^{-3}$ in the central part.
The displaced amount of gas is then expected to relax and sink toward the center 
between 2 successive outbursts. Or,
in a more realistic scenario, as gas is displaced (in a bi-conical nature), fresh gas can fall in perpendicular to the jet axis on the same timescale. This can justify our assumption
that the frequency of the AGN outbursts is equal to the frequency of the gas oscillations.
Physically, this assumption seems reasonable since the two times are
expected to be similar, as it is the accretion of core gas that drives the outbursts.
However, we cannot totally exclude the possibility that either the length of an outburst is longer than the gas response time or the frequency of the outbursts
is larger than the one used in our fiducial model. In both alternatives, the oscillations
would be damped and the mechanism proposed inefficient.  
To finish, dark matter haloes and galaxies are continuously accreting matter. The corresponding gas merging/infall are responsible to drive the AGN activity which is proved to be sporadic with 10\% duty cycle. Each outflow expels important amount of gas which becomes
a potential source of material for future BH accretion.
According to these physical mechanisms, the periodical quasar activity is expected to be maintained during a long period, probably over several Gyr, timescale required for our mechanism to operate.

Our astrophysical justification for this timescale is the following. Ellipticals most likely form by many minor mergers (Bournaud, Jog \& Combes \cite{bournaud}). The major merger rate is too low. These are inevitably distributed over a few Gyr. If the gas supply in  each merging event feeds the AGN 
as well as enhances star formation, we inevitably expect the AGN activity and growth to be spread out, e.g. with a duty cycle of 10\% over many dynamical times.
The downsizing of AGN activity (and associated growth) supports such prolonged activity for typical ellipticals, with only the most massive forming early by major mergers.

Moreover, it has been 
shown that the brightness profiles $I(r)$ of nearly all ellipticals could be well fitted
with S\'ersic $I \propto r^{1/n}$ of index $n$ (see for instance,
 Caon, Capaccioli \& D'Onofrio \cite{caon}). However, some recent studies have suggested
that most  giant ellipticals exhibit  ``missing  light'' at small radii  
 with respect to the inward extrapolation of the S\'ersic profile
(see Graham \cite{graham} or Ferrarese et al. \cite{ferrarese}).
 Such observations are believed to be the
consequences of dissipative galaxy formation mechanisms such as those involving AGN
 energy feedback. In particular, massive SMBH binary tidal scouring  to produce the observed mass deficit was
 advocated by Milosavljevi{\'c} \& Merritt (\cite{milosavljevic}). 
  We note here that AGN-induced gas bulk motions are equally likely,  if not more
 natural, as a ubiquitous accompaniment to the gas-rich phase of massive spheroid formation.
Our model assumes a substantial gas component
  not only during spheroid formation but also
over the next Gyr of AGN activity. This constitutes a serious issue for all CDM models since gas infall is predicted. But as mentioned above,
 gas is expelled and or/induce to form stars at high efficiency as infrered from ultra-luminous
infrared galaxies (ULIRGs) and submillimeter galaxies (SMGs). Wind are observed as well as
high-efficiency star formation in these objects and an
increasing frequency of AGN beyond z of 1. It is at least 20\% locally and given the different
duty cycles for star formation and AGN. The feedback is likely to always be important.

 Indeed, in  order to account for the correlation between SMBH mass and spheroid velocity dispersion, the period of SMBH growth by accretion (the case for accretion growth is powerfully argued by Yu \& Tremaine \cite{yu}) {\it must} be followed by gas expulsion associated with SMBH-driven outflows.  It is reasonable to argue that the expulsion phase is preceded by a phase when SMBH are subdominant with respect to the canonical relation but are still capable of driving bulk flows of the gas supply associated with sporadic, fueling-driven outbursts.
Quenching of star formation by AGN activity, widely considered to play a role in massive elliptical formation, provides a
plausible  means of maintaining the gas bulk motions over several dynamical times.
Since stars behave as a ``collisionless'' fluid similar to dark matter, one should expect that
AGN activities will affect not only the DM density profile  but also the star density profile, 
confirmed by our results. In this
case,  the brightness in the inner part should be altered, 
in good agreement with  observational studies such as 
 Graham (\cite{graham}) and Ferrarese et al. (\cite{ferrarese}).

Finally, we draw a parallel between giant galaxies and galaxy clusters. 
Indeed, Chandra observations
have revealed the  presence of cold fronts in many galaxy clusters (see for instance 
Markevitch et al. 2000, 2002; Vikhlinin, Markevitch \& Murray \cite{vikhlinin}; Mazzotta et al. \cite{mazzotta}).  
These cold fronts are supposed to be due to the presence of a dense, cold gas cloud
moving with respect to a hotter one. Some observations suggest that such cold clouds  have
 ``sloshing'' motions around the center of the cluster,  believed to be the
consequences of either  
a recent major merger event or  feedback processes from a central AGN. In fact, our
 model cannot be applied on such scales. Indeed, for a typical 
galaxy cluster of $10^{15}\,h^{-1}M_\odot$, the velocity dispersion of DM in the central part is
more than 1000 km/s. Then to obtain similar effects to our giant galaxy model, 
 about $10^3$ times more energy is required.
 It seems unrealistic that one or even a  few AGNs located near the center of the cluster 
can provide such  an amount of energy and induce  gas bulk motions in a  coherent way.
 However for massive galaxies, the situation is quite different. Not only are the energetics
 favourable, but the bipolar nature of the AGN outburst must inevitably exert dynamical
 feedback during the gas-rich phase of spheroid formation.

\begin{acknowledgements}
S. P. acknowledges support from a PPARC rolling grant.
It is a pleasure to thank the anonymous referee, 
as well as L. Ferrarese and A. Graham for their useful 
comments which have significantly improved this paper.
S. P. also wants to thank J. Magorrian, A. Eyre, G. Mamon, J.A. de Freitas Pacheco
and T. Sousbie for
interesting conversations.
\end{acknowledgements}

\end{document}